\documentstyle[preprint,prl,aps]{revtex}
\begin{document}
\draft

\title{Enhancement of Persistent Currents by the Coulomb Interaction}

\author{D. Yoshioka and H. Kato}
\address{Institute of Physics,
         College of Arts and Sciences,
         University of Tokyo, \\Komaba, Meguro-ku, Tokyo 153, Japan}

\date{\today}

\maketitle
\begin{abstract}
The persistent current in three-dimensional mesoscopic rings is
investigated numerically.
The model is tight-binding one with random site-energies and
interaction between electrons.
The Hartree-Fock approximation is adopted for the interaction between
electrons, and models with up to $3\times 4\times 40$ sites
are investigated.
It is shown that the long-range Coulomb interaction enhances the
persistent current considerably for rings with finite width.
It is also shown that long-rangedness of the  interaction is essential
for the enhancement.
Screening of the random site-energy is attributed to this enhancement.
\end{abstract}

\narrowtext

It has been predicted by  B\"uttiker {\em et al.} \cite{but} that
an equilibrium persistent current flows
in a mesoscopic normal-metal ring threaded by a magnetic flux.
This prediction has been experimentally confirmed\cite{exp1,exp2,exp3}.
Especially, the experiment on single Au ring\cite{exp2}
revealed that the size of the
persistent current observed was quite large.
Namely the observed current was $I = (0.3 \sim 2.0) I_0$,
where $I_0 = ev_{F}/L$
is of the order of the persistent current in an ideal one-dimensional
ring with $v_F$ being the Fermi velocity and $L$ being the
circumference of the ring.
The current is periodic in the magnetic flux $\phi$ piercing the
ring with period
$\phi_0=h/e$ as expected.
Thus the persistent current was predicted theoretically and confirmed
by the experiment.
However, this is not the whole story.
Quantitative understanding of the current has not been established.

It has been clarified by numerical\cite{mon1,mon2}
and theoretical\cite{altland} investigations
that theory without taking into account the
interaction between electrons can give current which is too small
compared to the experiments\cite{exp1,exp2}.
Inclusion of the mutual interaction into theory is hard and difficult
to assess the validity of the approximation employed, but
several groups\cite{ae,schmid}
have done such calculations, and obtained results that
sample averaged current is of the order of $(l_e/L)I_0$.
In this case the current is periodic in
$\phi_0/2$, because sample averaging kills the component with period
$\phi_0$.
This current is much smaller than $I_0$ in the experimental situation
of $L >> l_e$, where $l_e$ is the elastic mean free path.
In these theories the Coulomb interaction acts mainly to maintain the
local charge neutrality of the system.
Another theory by Kopietz\cite{kop} takes into account
long-range Coulomb interaction
energy associated with the charge fluctuation.
This theory gives the average persistent current of the order of
$0.1I_0$ with period $\phi_0/2$.
This value is large enough to explain the experiments\cite{exp1}.
However, since the fluctuation should be suppressed by the Coulomb
interaction itself to maintain the approximate local charge
neutrality, it is not yet certain that such large value survives
improvement of the theory.

On the other hand, numerical investigation is free from approximations,
and has a possibility to  give a clear answer to the origin of the
large persistent current.
However, unfortunately rigorous treatments are limited to quite small
system sizes.
Thus until now only small one-dimensional systems have been
investigated with negative answer to the role of the Coulomb
interaction\cite{ab,monc,gog}.
Namely, it has been clarified that in one-dimensional systems the
repulsive interaction between electrons suppress the persistent
current in situations relevant to the experiment.
We have also investigated the one-dimensional system with the Coulomb
interaction between electrons\cite{ky}.
Our results were the same as others.
However, we also found that the Hartree-Fock approximation
gives almost the same
results as those of the exact diagonalization.
The merit of the Hartree-Fock approximation is that we
can deal with much larger systems.
Thus we now apply our method to three-dimensional rings to
investigate the effect of the Coulomb interaction.

In the present work we have found that for the three-dimensional ring
the Coulomb interaction enhances the persistent current.
We have clarified conditions for the enhancement.
We have found that long-rangedness of the Coulomb interaction is
essential to this enhancement:
Short-ranged interaction does not enhance the current.
We also found a condition for the cross section of the ring to
have the enhancement.

We consider the Anderson model with Coulomb interaction
between electrons.
For simplicity we neglect spin freedom of electrons.
Thus we consider a tight binding model on a cubic lattice.
Our model ring consists of $N_\ell \times N_w \times N_h$ sites:
circumference $N_\ell a$, width $N_w a$, and
height $N_ha$, where $a$ is the lattice constant.
The coordinate of the $i$-th site is given by three integers
$(\ell_i,w_i,h_i)$,
where $1 \leq \ell_i \leq N_\ell$, $1 \leq w_i \leq N_w$, and
$1 \leq h_i \leq N_h$.
Each site has random site-energy $\varepsilon_i$, which
has uniform distribution between $-W/2$ and $W/2$.
Thus our Hamiltonian is
\begin{eqnarray}
\label{ex}
{\cal H}&=&-\sum_{i,j}( t_{i,j} e^{i \theta_{i,j}} c_i^{\dagger} c_j
        + h.c.)
        + \sum_i \varepsilon_i c_i^{\dagger} c_i \nonumber \\
    & & + \frac12 \sum_{i \ne j} V \frac{a}{r_{ij}}
       (c_i^\dagger c_i - \rho_0)(c_j^\dagger c_j - \rho_0),
\end{eqnarray}
where $c_i^\dagger (c_i)$ is the creation (destruction) operator of
a spinless electron,
$t_{i,j}$ is non-zero and equal to $t$ only for nearest neighbor
hopping, $\theta_{i,j}$ is non-zero only for hopping along the
circumference of the ring and has a value $2\pi\phi/N_\ell \phi_0$,
$V=e^2/4\pi\epsilon a$ is the strength of the Coulomb interaction,
and $\rho_0$ is neutralizing positive charge at each site.
To make a ring which has the same lattice constant everywhere we
have embedded
our three-dimensional ring in the four-dimensional space, so
the distance $r_{ij}$ between $i$, $j$ sites  is given by
\begin{eqnarray}
\label{dist}
(\frac{r_{ij}}{a})^2 &=& 2R^2\{1 - \cos[\frac{2\pi}{N_\ell}
(\ell_i - \ell_j)]\} \nonumber \\
& &+ (w_i - w_j)^2 + (h_i - h_j)^2,
\end{eqnarray}
where $R=[2\sin(\pi/N_\ell)]^{-1}$ is the radius of the ring.

The following treatment is the same as before\cite{ky}:
The interaction term is decoupled following standard Hartree-Fock
scheme by introducing the average $\langle c_i^\dagger c_j \rangle$,
which is determined self-consistently by iteration.
Once the ground state energy $E_g(\phi)$ is determined
self-consistently, the persistent
current $I(\phi)$ is given by the derivative of the energy
$E_g(\phi)$ with respect to the magnetic flux:
\begin{eqnarray}
\label{pc}
I(\phi)=-\frac{\partial E_g(\phi)}{\partial \phi},
\end{eqnarray}
Since $E_g(\phi)$ is an even function of $\phi$ and periodic with
period $\phi_0$,
we first Fourier analyze the energy,
\begin{eqnarray}
\label{ef}
E(\phi) = E_0+ \sum_{n=1}^\infty E_n
\cos(2\pi n \frac{\phi}{\phi_0}),
\end{eqnarray}
and obtain the Fourier components of the current:
\begin{eqnarray}
\label{cf}
I(\phi) &=& \sum_{n=1}^\infty \frac{2\pi n}{\phi_0} E_n
         \sin(2\pi n\frac{\phi}{\phi_0}) \nonumber \\
       &\equiv&\sum_{n=1}^{\infty} I_n \frac{t}{\phi_0}
        \sin(2\pi n\frac{\phi}{\phi_0}).
\end{eqnarray}
Here $I_n$ is a dimensionless number, which depends on realization of
the random site-energy and the strength of the Coulomb interaction.
In the following we mainly concentrate on the $\phi_0$-periodic
component, $I_1$.

We have investigated up to 480-site systems by this method, the
largest being $N_\ell = 40$, $N_w=4$, and $N_h=3$ with 200 electrons.
The randomness is chosen such that the sample is in the diffusive
regime,
which is realized around $W/t= 2$ for the samples we have considered.
The results for the largest four samples are shown in Figs.1, 2 and 3.
First in Fig.1 dependence on the randomness is shown for the
{\it non-interacting} case to clarify the situation before we take
the interaction into account.
An arrow indicates the value of $I_1=4\pi/N_\ell$.
The size of this current $(4\pi/N_\ell)t\phi_0$ is equal
to $ev_F/L = I_0$,
if we use $v_F = 2|t|a/\hbar$;
namely it indicates the size of the persistent current for a
pure one-dimensional ring of the same circumference.
Figure 2 shows how the interaction enhances the current.
It is seen that the interaction enhances the current beyond the
value of the pure 1-d system, $I_0$.
On the other hand Fig.3 shows the results obtained by replacing the
long-range Coulomb interaction by short ranged one,
namely $V\times (a/r_{ij})$ is replaced by $V\delta(r_{ij} - a)$.
It has the same value for the nearest neighbor but vanishes otherwise.
The results show essentially no enhancement for this case.
This comparison clarifies the origin of the enhancement:
long-rangedness is essential for the enhancement.
Quite similar results have been obtained for smaller samples
($33\times 3\times 2$ sites and $16\times 3\times 2$ sites).
It should be noticed that the sign of $I_1$ is not fixed:
In these figures the positive values are shown by open symbols and
negative ones are shown by closed symbols.
It can be both positive (paramagnetic) or negative (diamagnetic)
depending on samples.

Since our present result is opposite to that of one-dimensional
system,
we investigated the boundary between the
enhancement and the suppression.
Both two-dimensional and tree-dimensional rings were examined.
For the two-dimensional system, $N_w$ was kept to unity and $N_h$ was
increased from one to six.
Definite enhancement by $V/t$ was observed for systems with
$N_h \geq 4$.
For the three-dimensional system, $N_w$ was kept to
two and $N_h$ is increased from one to three.
It was found that $N_h\geq 2$ is enough to have the enhancement.
Thus in both cases enhancement occurs for systems
with $N_w \times N_h \geq 4$; otherwise the current is suppressed.

In the present work we have shown for the first time that the Coulomb
interaction can enhance the persistent current beyond the value
for an ideal one-dimensional ring, $I_0$\cite{mg2}.
Then the next question will be if the parameters used here are
relevant to the experimental situation or not.
We think that choice of the randomness $W/t$ poses no question.
We need to be in the diffusive regime, and within this regime
$V/t$ dependence is  qualitatively similar.
Appropriate choice of $V/t$ is a difficult question,
since size of our system is still too small
compared to the experiment, where about $10^{8}$ atoms are involved.
Here we only comment that if our system is actually made of Au atoms,
$t$ is of the order of 1eV, and bare Coulomb energy between one
lattice constant is larger than $t$: $V = e^2/4\pi \epsilon a
\simeq 5eV > t$.
So even if we consider the optical dielectric constant
$\epsilon(\infty)$
by ion core, $V$ and $t$ are of the same order.
We have found strong enhancement even when $V$ is one order of
magnitude smaller than $t$.
In the actual system the interaction is screened by electrons,
but such screening effect is included in our self-consistent
solution.
We need the flux dependence of the total ground state energy
which includes the energy associated with the screening.
Therefore use of the long-range Coulomb interaction is needed and
justified.

Our treatment is not exact as we use the Hartree-Fock approximation.
We justify it in two ways.
Firstly, as shown in ref.\cite{ky} the results by our Hartree-Fock
approximation is consistent with those by the exact diagonalization
for small one-dimensional systems.
Secondly, in this work we have clarified that the long-range part of
the interaction is essential for the enhancement.
It is expected that correlation effects beyond the Hartree-Fock
approximation mainly affects the short-range behavior.
However, since we consider spinless electrons, main part of the
short-range correlation, on-site correlation, has already been taken
into account.
Thus we expect that our present result is almost the same as what
would be obtained if the correlation were taken into account exactly.

The final question will be the mechanism of the enhancement.
One possible explanation is the screening of the random potential.
The electron at the Fermi level feels screened site-energy which is
flattened, so the current
is enhanced.
In the case of one-dimensional system the screening is also effective,
but electrons are not allowed to pass through each other, thus
the current is suppressed by localization of some of electrons.
When the cross section of the ring is small enough
(e.g. $N_w\times N_h < 4$),
this blockade is still effective.
This explanation is justified as follows.
We have examined which term in the Hamiltonian mainly contribute to the
current.
We found that the contribution from the first two terms of
the Hamiltonian Eq.\ (\ref{ex}), i.e. the single-particle terms,
$I_s(\phi) = - (\partial/\partial\phi)\langle -\sum_{i \ne j}
(t_{i,j}e^{i\theta_{ij}}c_i^\dagger c_j + h.c.)
 + \sum_i \varepsilon_i
c_i^\dagger c_i \rangle$,
amounts to about 90\% of the total current,
where $\langle ~~ \rangle$ means expectation values with respect to
the self-consistent ground state.
The remaining 10\% of the current comes from the interaction term
of the Hamiltonian which consists of the Hartree contribution,
$I_H(\phi) = -(\partial/\partial\phi) \sum_{i \ne j}
V(a/2r_{ij})\langle c_i^\dagger c_i\rangle
\langle c_j^\dagger c_j\rangle$,
and the Fock contribution,
$I_F(\phi) = (\partial/\partial\phi) \sum_{i \ne j}
V(a/2r_{ij})\langle c_i^\dagger c_j\rangle
 \langle c_j^\dagger c_i\rangle$,
the latter being usually smaller than the former.
Thus the enhancement is caused mainly by the single particle part of
the Hamiltonian.
This means that the ground state is quite different from that of the
non-interacting system,
where we get quite small persistent current.
We expect that the electron states near the Fermi level are
more like those of the system with $W=V=0$.
This explains why the theories using the diagrammatic
expansion\cite{ae,schmid,g2,alt} could not get the enhancement:
Since the ground state has changed considerably from that without
the Coulomb interaction, perturbational
calculation cannot give correct answer.

In conclusion we have shown for the first time that the persistent
current is enhanced beyond the ideal one-dimensional value by the
long-range Coulomb interaction for three-dimensional rings.
We have clarified that a ring with finite cross section
and long-rangedness of the interaction are necessary for the
enhancement.
We attribute the enhancement to the screening of the
random site-energies.

One of the authors (D. Y.) thanks Aspen Center for Physics
where part of this work was done.
The numerical calculation was done by HITAC S-3800/480 at the
Computer Center of the University of Tokyo.


%
%
\begin{figure}
\caption{Fourier component of the persistent current with period
$\phi_0$ is plotted as a function of the site randomness $W/t$ for
the non-interacting system ($V/t=0$).  The four different symbols,
circle, two kinds of triangles,
and square, indicate four different realization of the randomness.
The absolute values of the current are plotted with the open
(closed) symbols indicating that the value is positive (negative).
An arrow with letter $I_0$ indicates the value of
$I_1$ where $I_1t/\phi_0 = I_0$.}
\label{fig:w-dep}
\end{figure}

\begin{figure}
\caption{Fourier component of the persistent current with period
$\phi_0$ is plotted as a function of the strength of the Coulomb
interaction $V/t$ for the randomness $W/t=2.5$.
The same four samples in Fig.1 are used with the same symbols.
Here also the open (closed) symbols mean that the value is positive
(negative).
An arrow with letter $I_0$ indicates the value of $I_1$ where
$I_1t/\phi_0 = I_0$.}
\label{fig:v-dep}
\end{figure}

\begin{figure}
\caption{Fourier component of the persistent current with period
$\phi_0$ is plotted as a function of the strength of the short-range
interaction $V/t$ for the randomness $W/t=2.5$.
The same four samples in Figs.1 and 2 are used with the same symbols.
Here also the open (closed) symbols mean that
the value is positive (negative).}
\label{fig:sr}
\end{figure}

\end{document}